\documentclass[sigconf]{acmart}

\usepackage{booktabs} 
\usepackage{color}

\setcopyright{acmcopyright}

\acmDOI{10.475/123_4}

\acmISBN{123-4567-24-567/08/06}

\acmConference{Communications of the ACM}{May 2018}{New York, NY, USA} 

\acmYear{2018}
\copyrightyear{2018}
\acmArticle{4}
\acmPrice{15.00}

\begin{document}
\title{SONYC: A System for the Monitoring, Analysis and Mitigation of Urban Noise Pollution}

\author{Juan P. Bello}
\affiliation{%
  \institution{New York University}
  \city{New York} 
  \state{NY} 
  \postcode{10012}
}
\email{jpbello@nyu.edu}

\author{Claudio Silva}
\affiliation{%
  \institution{New York University}
  \city{New York} 
  \state{NY} 
  \postcode{10012}
}
\email{csilva@nyu.edu}

\author{Oded Nov}
\affiliation{%
  \institution{New York University}
  \city{New York} 
  \state{NY} 
  \postcode{10012}
}
\email{onov@nyu.edu}

\author{R. Luke DuBois}
\affiliation{%
  \institution{New York University}
  \city{New York} 
  \state{NY} 
  \postcode{10012}
}
\email{dubois@nyu.edu}

\author{Anish Arora}
\affiliation{%
  \institution{Ohio State University}
  \city{Columbus} 
  \state{OH} 
  \postcode{43210}
}
\email{arora.9@osu.edu}

\author{Justin Salamon}
\affiliation{%
  \institution{New York University}
  \city{New York} 
  \state{NY} 
  \postcode{10012}
}
\email{justin.salamon@nyu.edu}

\author{Charles Mydlarz}
\affiliation{%
  \institution{New York University}
  \city{New York} 
  \state{NY} 
  \postcode{10012}
}
\email{cmydlarz@nyu.edu}

\author{Harish Doraiswamy}
\affiliation{%
  \institution{New York University}
  \city{New York} 
  \state{NY} 
  \postcode{10012}
}
\email{harishd@nyu.edu}

\renewcommand{\shortauthors}{J.P. Bello et al.}



%
%



\maketitle

\section{Noise Pollution}

Noise refers to unwanted or harmful sound from environmental sources such as traffic, construction, 
industrial and social activities. 
Noise pollution is one of the topmost quality of life issues for urban residents in the United States, with over 70 million people across the country exposed to noise levels beyond the limit of what the EPA considers to be harmful \cite{hammer2014}. 

Such levels of exposure have proven effects on health, 
including acute effects such as sleep disruption, 
and 
long-term effects 
such as hypertension, heart disease and hearing loss \cite{bronzaft2007neighborhood, hammer2014, fritschi2012introduction}.
In addition, there is evidence of impact on educational performance, with studies showing that noise pollution produces learning and cognitive impairment in children, resulting 
in decreased memory, reading skills and lower test scores 
\cite{bronzaft2007neighborhood, basner2014auditory}. 

The economic impact of noise is also significant. The WHO 
estimates that, in Western Europe alone, 1 million healthy life-years are lost annually to environmental noise \cite{fritschi2012introduction}. 
Other estimates put the external cost of noise-related health issues in the EU between 0.3-0.4\% of GDP \cite{cedelft2008}, 0.2\% in Japan \cite{mizutani2011}. 
Studies in the US and Europe also demonstrate 
the relationship between environmental noise and real state markets, with housing prices falling as much as 2\% per decibel (dB) of noise increase 
\cite{nelson1982,theebe2004}.

In short, 
noise pollution is not merely an annoyance but an important problem with broad-ranging societal effects that apply, to a varying extent, to a significant portion of the population. 
Therefore, it is clear that effective noise mitigation is in the public interest, with proven health, economic, and quality-of-life impact.

\section{The Challenge of Noise Mitigation}
Noise can be mitigated at the receiver's end, e.g. by wearing ear plugs, or along the transmission path, e.g. by erecting sound barriers around major roads. These strategies do not reduce noise emissions, and 
place the burden of mitigation on the receiver \cite{hammer2014}. Alternatively, we can mitigate noise at the source, e.g. by 
designing aircrafts with quieter engines, acoustically treating clubs, 
using muffled jackhammers in roadworks, 
or stopping unnecessary honking. 
Such actions are commonly incentivized with a regulatory framework that uses fines and other penalties to raise the cost of emitting noise 
\cite{quieteramerica}. However, enforcing noise codes in large urban areas, to the point where they effectively deter noise emissions, is far from trivial.

Take NYC as an example. 
Beyond occasional physical inspections, 
the city monitors noise via its 311 service for civil complaints. 
Since 2010, 311 has logged more than 2.3 million noise-related complaints, significantly more than for any other issue~\footnote{http://www1.nyc.gov/311}. 
This averages about 834 complaints a day, the largest citizen noise reporting system anywhere in the world. 
However, research by NYC's Department of Health and Mental Hygiene (DOHMH) shows that 311 data does not accurately capture information about all noise exposure in the city \cite{EpiBrief2014}. 
Their study 
shows the top sources of disruptive noise to be traffic, sirens and construction; the effect to be similar in the boroughs of Manhattan, Brooklyn and the Bronx; and low-income and unemployed New Yorkers 
 amongst those most frequently exposed. 
In contrast, 311 noise complaint data collected for the same period 
prioritizes social noise such as parties, car alarms, loud talking, music and TV, 
with a minority of 
complaints citing traffic or construction. 
Notably, residents of Manhattan, where most affluent New Yorkers live, are more than twice as likely to file 311 complaints than those in other boroughs. This clearly highlights the need for collecting objective measurements of noise across the city, alongside citizen reporting, to fully characterize the phenomenon.

A closely related issue concerns how to effectively respond to potential violations of the noise code. In NYC, the subset of noise complaints pertaining to static, systemic sources such as construction, animals, traffic, air conditioning and ventilation units, are routed to the Department of Environmental Protection (DEP). The DEP employs about 50 highly-qualified inspectors to measure sound levels and issue a notice of violation whenever needed. Unfortunately, the limited human resources and the high volume of complaints result in average response times of more than 5 days. Given the transient nature of sound, a very small proportion of inspections actually result in a violation observed, let alone penalized.

To make matters worse, even when noise sources are active during inspections, it is hard to isolate their effect. Noise is commonly measured in overall sound pressure levels (SPL) expressed in A-weighted decibels (dBA) 
\cite{quieteramerica}, which aggregate all sound energy in an acoustic scene. Thus, existing technologies cannot isolate the effect of offending sources, especially in urban environments featuring a large number of sounds. As a result, inspectors have to resort to long and complicated measurement strategies that often require help from the people responsible for the violation in the first place -- an additional factor contributing to the difficulty and reduced efficiency of the enforcement process. 

\smallskip

In this paper we outline the opportunities and challenges associated with SONYC, a cyber-physical systems approach to the monitoring, analysis and mitigation of urban noise pollution. This initiative connects multiple sub-fields of computing including wireless sensor networks, machine learning, collaborative and social computing and computer graphics, to create a potentially transformative solution to this important quality-of-life issue. To illustrate this potential, we present findings from an initial study showing how SONYC can help understand and address important gaps in the process of urban noise mitigation.

\section{SONYC}

\begin{figure*}[t]
\begin{center}
\includegraphics[width=0.7\textwidth]{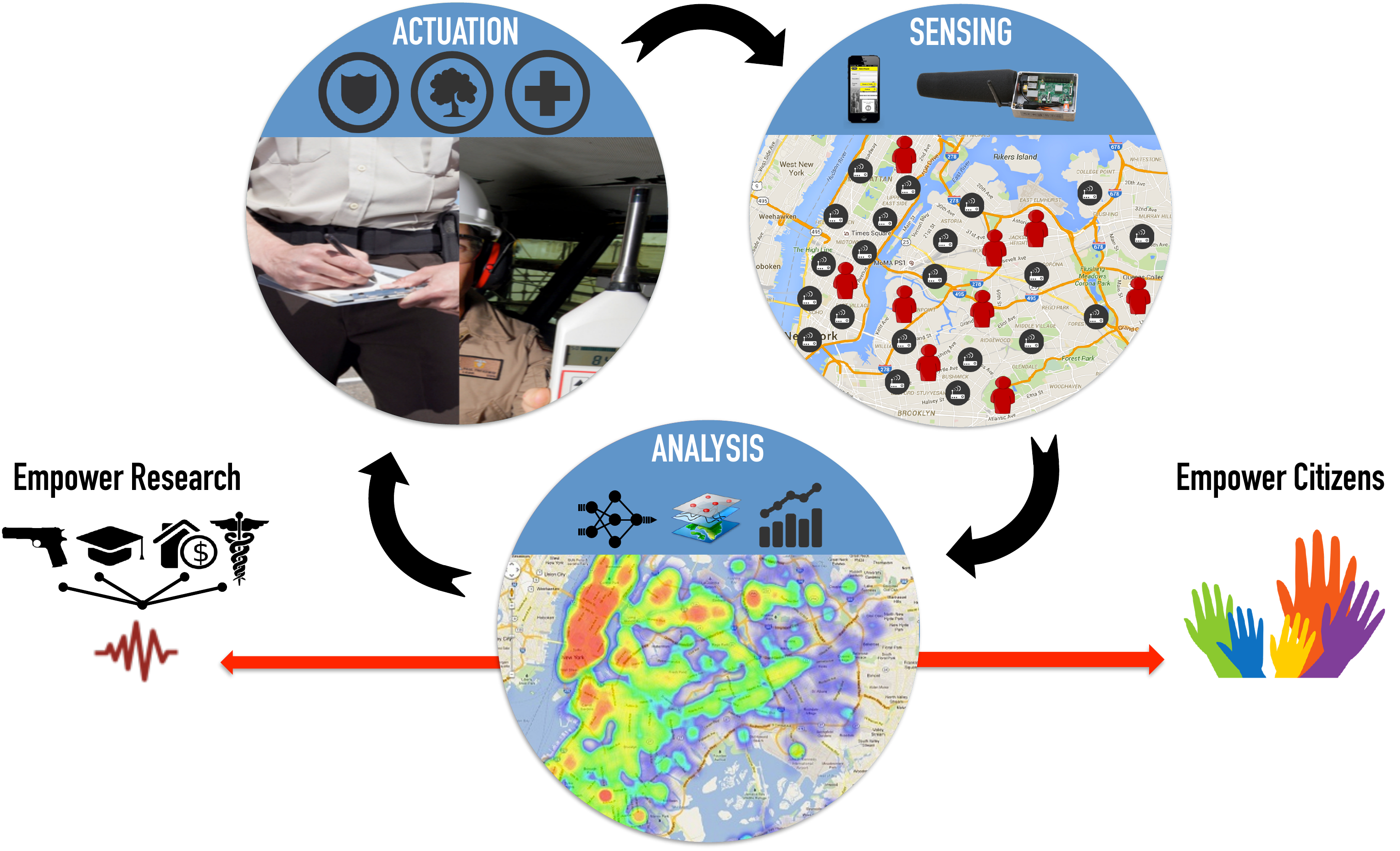}
\end{center}
\caption{The SONYC cyber-physical system loop including intelligent sensing, noise analysis at city-scale and data-driven mitigation. SONYC empowers new research in the social sciences and public health, while empowering citizens to improve their communities.} 
\label{fig:SONYC_CPS}
\end{figure*}

There have been multiple attempts to inject technological solutions to improve the cycle of urban noise pollution. For example, various initiatives have used mobile devices to crowdsource instantaneous SPL measurements, noise labels and subjective responses \cite{ruge2013soundofthecity,schweizer2012noisemap,becker2013awareness}, but they lag well behind the coverage in space-time of civic complaint systems such as 311, while the reliability of their objective measurements suffers from lack of adequate calibration. Others have deployed static sensing solutions that are often too costly to scale up and fail to go beyond the capabilities of standard noise meters \cite{pham2013streaming,bell2013novel,steele_2013}. On the analytical side, there has been a significant amount of work on noise maps generated from sound propagation models for major urban noise sources such as industry, road, rail and air traffic \cite{bandkprediction_2015,kaliski2007community}. However, these maps lack temporal dynamics and make modeling assumptions that often render them too inaccurate to support mitigation or action planning \cite{ausejo2010study}. Very few of these initiatives have involved acting upon the sensed or modeled data to affect noise emissions, and even fewer have counted with the participation of local governments \cite{Manvell_SADMAM04}.

{\bf SONYC} (Sounds of New York City), the novel solution depicted in Fig.~\ref{fig:SONYC_CPS}, aims to address these limitations via an integrated cyber-physical systems' approach to noise pollution. 

First, it proposes a low-cost, intelligent \emph{sensing} platform capable of continuous, real-time, accurate and source-specific noise monitoring. Our sensing solution is scalable in terms of coverage and power consumption, does not suffer from the same biases as 311-style reporting, and goes well beyond SPL-based measurements of the acoustic environment.

Second, SONYC adds new layers of cutting-edge data science methods for large-scale noise \emph{analysis}. These include predictive noise modeling in off-network locations using spatial statistics and physical modeling, the development of interactive 3D visualizations of noise activity across time and space to enable a better understanding of noise patterns, and novel information retrieval tools that exploit the topology of noise events to facilitate search and discovery.

Third, this sensing and analysis framework is used to improve {\it mitigation} in two ways: first, by enabling optimized, data-driven planning and scheduling of inspections by the local government, thus improving the likelihood that code violations will be detected and enforced; and second, by increasing the flow of information to those in a position to control emissions -- e.g. building and construction-site managers, drivers, neighbors -- thus providing credible incentives for self-regulation. Because the system constantly monitors and analyzes noise pollution, it generates information that can be used to validate, and iteratively refine any noise mitigating strategy.

Take for example a scenario in which the system integrates information from the sensor network and 311 to identify a pattern of after-hours jackhammer activity around a construction site. This information triggers targeted inspections by the DEP which results in the issuing of a violation. Through statistical analysis we can then validate whether the action is short-lived in time, or whether its effect propagates to neighboring construction sites or distant ones 
by the same company. 
By systematically monitoring interventions, we can understand how often penalties need to be imparted before the effect becomes long-term. 
The overarching goal is to understand how to minimize the cost of interventions while maximizing noise mitigation, a classic resource allocation problem that motivates much research on smart-cities initiatives. 

All of this is made possible by formulating our solution in terms of a cyber-physical system (CPS). However, unlike most CPS work in the literature, the distributed and decentralized nature of the problem requires the leveraging of multiple socio-economic incentives -- e.g. fines or peer comparisons -- to exercise indirect control on tens of thousands of sub-systems contributing noise emissions. It also calls for the development and implementation of a set of novel mechanisms for integrating humans in the CPS loop at scale and at multiple levels of the system's management hierarchy, including the extensive use of human-computer interaction (HCI) research in, e.g., citizen science and data visualization, to facilitate seamless interactions between humans and cyber-infrastructure. It is worth emphasizing that this line of work is fundamentally different from current research on human-in-the-loop CPS, which is often focused on applications where control is centralized and fully or mostly automated, while there is usually a single human involved -- e.g. in assistive robots and intelligent prosthetics. We believe that the synthesis of approaches from social computing, citizen science and data science to advance the integration, management and control of large and variable numbers of human agents in CPS is potentially transformative, addressing a crucial bottleneck for the widespread adoption of similar methods in socio-technical systems such as transportation networks, power grids, smart buildings, environmental control and smart cities.

Finally, SONYC uses NYC, the largest, densest, noisiest city in North America, as its experimental ground. 
The city has long been at the forefront of discussions about noise pollution, has an 
exemplary noise code \footnote{http://www.nyc.gov/html/dep/html/noise/index.shtml}  and, 
in 311, the largest citizen noise reporting system. 
Beyond noise, NYC collects large amounts of data about everything from public safety, traffic, taxi activity, and construction, and makes much of that information open
\footnote{https://nycopendata.socrata.com}. 
Our work involves close collaboration with city agencies such as the Department of Environmental Protection (DEP), the Department of Health and Mental Hygiene (DOHMH), various business improvement districts (BIDs), and private initiatives, such as LinkNYC, that provide access to existing infrastructure. Thus, as a powerful sensing and analysis infrastructure, SONYC holds the potential to empower new research in environmental psychology, public health and public policy, as well as empower citizens seeking to improve their own communities.

\smallskip

In the following sections we describe the technology and methods underpinning the project, present some of our early findings, and discuss future challenges.  

\section{Acoustic Sensor Network}

\begin{figure*}[t]
\begin{center}
\includegraphics[width=0.65\textwidth]{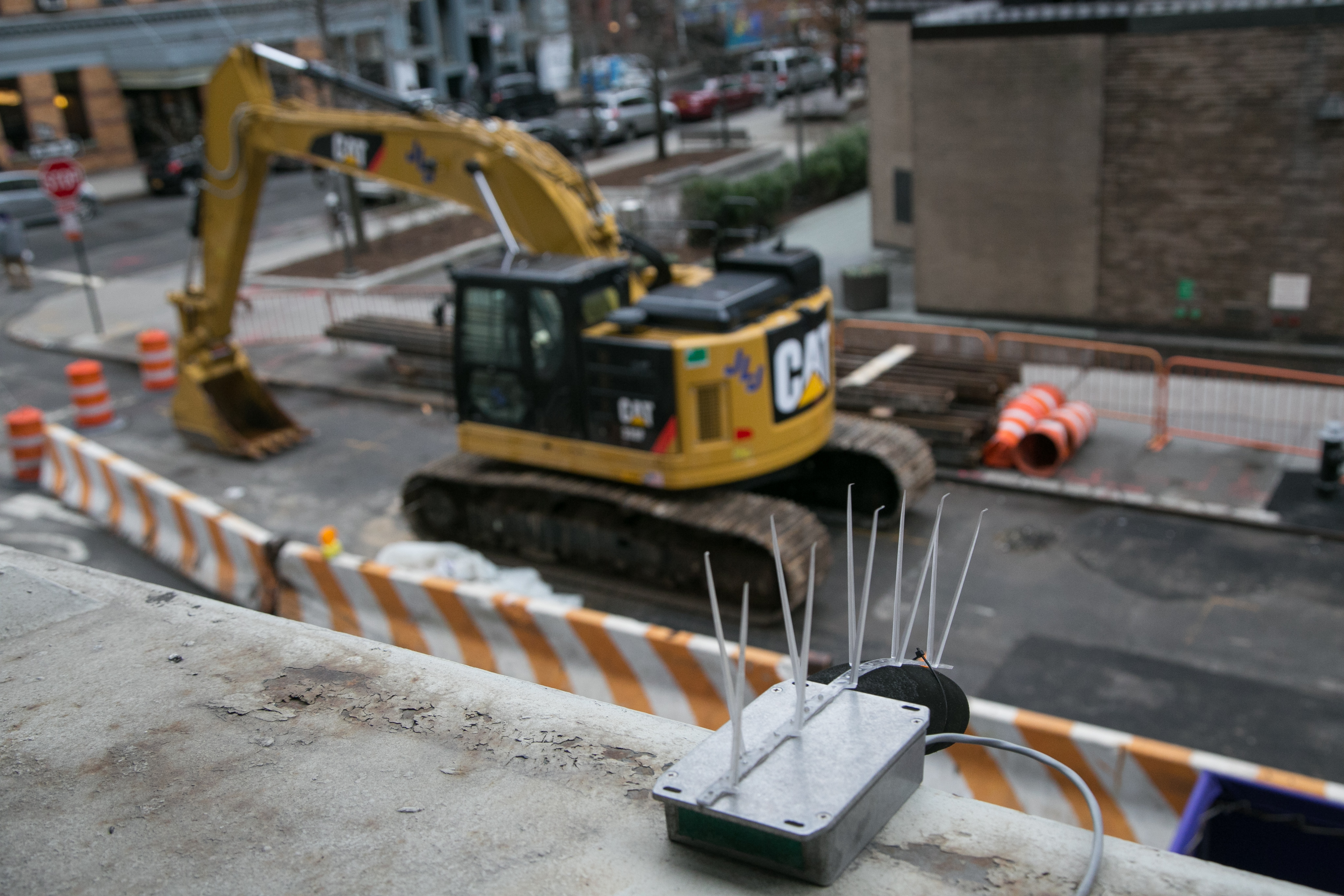}
\end{center}
\caption{Acoustic sensing unit deployed on a New York City street.}
\label{fig:sensor}
\end{figure*}

As discussed earlier, SONYC's intelligent sensing platform should be scalable and capable of source identification and high-quality, 24/7 noise monitoring. 

To that end we have developed an acoustic sensor \cite{Mydlarz_AppliedAcoustics17} -- shown in Figure \ref{fig:sensor} -- based on the popular Raspberry Pi single-board computer (SBC) outfitted with a custom microelectromechanical systems (MEMS) microphone module. MEMS microphones are chosen for their low cost, consistency across units and size, which can be 10x smaller than traditional microphones. Our custom, standalone microphone module includes additional circuitry such as in-house ADC and pre-amps stages, as well as an on-board micro-controller which enables pre-processing of the incoming audio signal to compensate for the microphone's frequency response. 
The digital MEMS microphone features a wide dynamic range of 32-120\,dBA, ensuring all urban sound pressure levels can be effectively monitored. It was calibrated using a precision grade sound-level meter as reference under low-noise, anechoic conditions, and was empirically shown to produce sound pressure level data at an accuracy compliant with the ANSI Type-2 standard \cite{quieteramerica} that is required by most local and national noise codes. 

The sensor's computing core is housed within an aluminum casing chosen to reduce RFI interference and solar heat gain. The microphone module is mounted externally via a repositionable metal goose-neck allowing the sensor node to be reconfigured for deployment in varying locations such as building sides, light poles and building ledges. Apart from continuous SPL measurements, the nodes will be sampling 10-second audio snippets at random intervals during a limited period of time. This is to collect data to train and benchmark our machine listening solutions. The audio is compressed using the lossless FLAC encoder, and encrypted using 4096\,bit AES encryption and the RSA public/private key-pair encryption algorithm. 
Sensor nodes communicate with the server via a Virtual Private Network (VPN), uploading audio and SPL data at 1 minute intervals. 

At the time of writing, the cost in parts of each sensor is around US\$80 using mostly off-the-shelf components. We fully expect the unit cost to be significantly reduced via custom redesigns for high-volume, third-party assembly. However, even at the current price tag, SONYC sensors are significantly more affordable, and thus amenable to large-scale deployment, than existing noise monitoring solutions. 
Furthermore, this increase in affordability does not come at the expense of measurement accuracy, with our sensors performing comparably to high-quality devices that are orders of magnitude more costly, while outperforming solutions in the same price range. Finally, the addition of a powerful and dedicated computing core, opens the possibility for edge computing, particularly for in-situ machine listening intended to automatically and robustly identify the presence of common sound sources. This is a unique feature of SONYC that is well beyond the capabilities of existing noise monitoring solutions.

\section{Machine Listening on the Edge}

Machine listening is the auditory counterpart to computer vision, combining techniques from signal processing and machine learning to develop systems 
able to extract meaningful information from sounds. In the context of SONYC, we are focused on developing computational methods to detect specific types of sound sources, such as jackhammers, idling engines, car horns, or police sirens, automatically from environmental audio. This is a challenging problem given the complexity and diversity of sources, auditory scenes and background conditions that can be found in urban acoustic environments. 

To address these challenges we have contributed an urban sound taxonomy, annotated datasets, and various cutting-edge methods for urban sound source identification \cite{Salamon:UrbanSound:ACMMM:14,Salamon_SPL17}. Our research shows that feature learning, even using simple dictionary-based methods such as spherical k-means, makes for significant improvement in performance upon the traditional approach of feature engineering. Importantly, we have found that temporal shift invariance, whether using modulation spectra or deep convolutional networks, is crucial not only for overall higher accuracy, but also to increase robustness in low signal-to-noise (SNR) conditions, as is the case when sources of interest are in the background of acoustic scenes. Shift invariance also results in more compact machines that can be trained with less data, thus adding value for edge computing. More recent results highlight the benefits of using convolutional-recurrent architectures, as well as ensembles of various models via late fusion. 

Deep learning models necessitate high volumes of labeled data, which have been traditionally unavailable for environmental sound. To alleviate this, we have developed an audio data augmentation framework, that systematically deforms the data using well-known audio transformations such as time stretching, pitch shifting, dynamic range compression and the addition of background noise at different SNRs, significantly increasing the amount of data available for model training. We have additionally developed an open-source tool for soundscape synthesis \cite{Salamon_WASPAA17}. Given a collection of isolated sound events, this tool acts as a high-level sequencer that can generate multiple soundscapes from a single, probabilistically defined, ``specification''. We have utilized this method to generate large datasets of perfectly annotated data in order to assess algorithmic performance as a function of, e.g., maximum polyphony and SNR ratio, studies that would be prohibitive at this scale and precision using manually-annotated data.

The combination of an augmented training set and the increased capacity and representational power of deep learning models results in state-of-the-art performance. Our current models can perform robust 10-class, multi-label classification in real-time, running on a laptop machine, and will soon be adapted to run under the computational constraints of the Raspberry Pi. 

However, despite the advantages of data augmentation and synthesis, the lack of significant amounts of annotated data for supervised learning remain the main bottleneck. To address this need, we developed a framework for web-based human audio annotation, and conducted a large-scale, experimental study on how visualization aids and acoustic conditions affect the annotation process and efectiveness \cite{Cartwright_CSCW18}. In this work, we aimed to quantify the reliability/redundancy trade-off in crowdsourced soundscape annotation, investigate how visualizations affect accuracy and efficiency, and characterize how performance varies as a function of audio characteristics. Our study followed a between-subjects factorial experimental design in which we tested 18 different experimental with 
540 participants recruited via Amazon's Mechanical Turk.

We found that more complex audio scenes result in lower annotator agreement, and that spectrogram visualizations are superior in producing higher quality annotations at lower cost of time and human labor. Given enough time, all tested visualization aids enable annotators to identify sound
events with similar recall, but the spectrogram visualization
enables annotators to identify sounds more quickly. We speculate that this may be because annotators are able to more easily identify visual patterns in the spectrogram, which in turn enables them to identify sound events and their boundaries in time more precisely and efficiently. We also see that participants learn how to use each interface more effectively over time, suggesting that we can expect higher quality annotations with even a small amount of additional training.

Crucially, we found that the value of additional annotators decreased after 5-10 annotators and that 16 annotators captured 90\%  of the gain in annotation quality. However, when resources are limited and cost is a concern, our findings suggest that five annotators may be a reasonable choice for reliable annotation with respect to the trade-off between cost and quality. These findings are valuable for the design of audio annotation interfaces, and the use of crowdsourcing and citizen science strategies for audio annotation at scale.

\section{Noise Analytics} 

One of the main promises of SONYC is the ability to analyze and understand noise pollution at city-scale in an interactive and efficient manner. At the time of writing, we have already deployed 45 sensors, primarily in the Greenwich Village neighborhood of NYC, but also in other locations in Manhattan and Brooklyn. Collectively, these sensors have gathered the equivalent of 8 years of audio data, more than twice that in sound pressure levels and telemetry. 
These numbers give a clear indication of the magnitude of the problem from a data analytics perspective.

Therefore we are developing a flexible and powerful visual analytics framework that can enable the visualization of noise levels in the context of the city together with other related urban data streams. Working with urban data poses new research challenges.  Although there has been much work on scaling databases for big data, existing technologies do not meet the requirements needed to interactively
explore massive or even reasonably-sized data sets~\cite{fekete@deb2012}. Accomplishing interactivity not only
requires efficient techniques for data and query management, it
also raises the need for scalable visualization techniques capable of
rendering large amount of information. In addition, visualizations and
interfaces must be easily understood by both domain experts and
non-expert users, including crowdsourcing workers and volunteers, and bear meaningful relationship to the properties of the data in the physical world, which in the case of sound implies the need for 3-dimensional visualization.

\begin{figure*}
\centering
\includegraphics[width=0.75\linewidth]{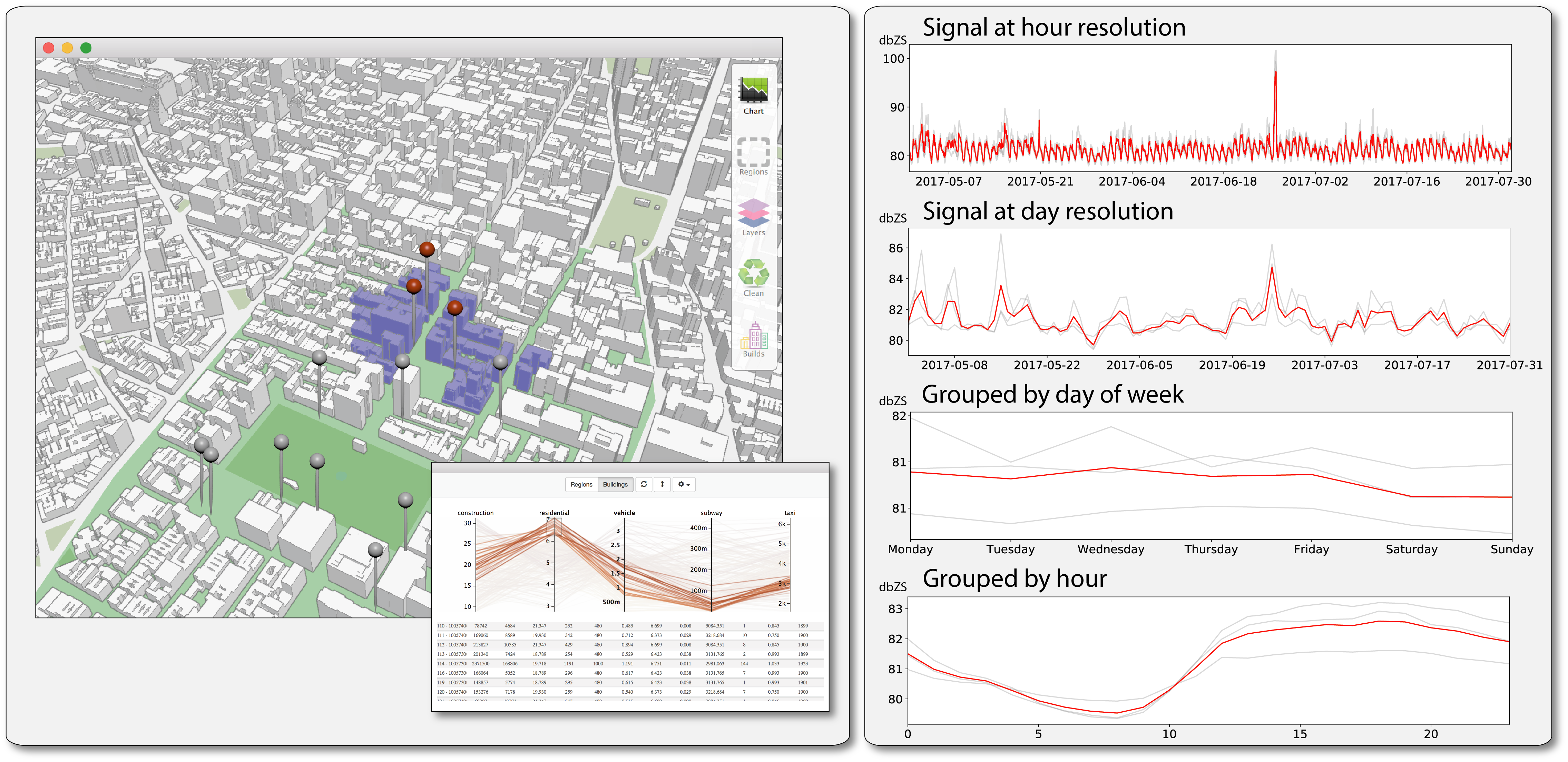}
\caption{(left) An interactive 3D visualization of a NYC neighborhood using \emph{Urbane}. By selecting specific sensors (red pins) and buildings (purple) we can easily and efficiently retrieve and visualize multiple data streams associated with those locations. (right) SPL data at various resolutions and time scales retrieved using the \emph{time lattice}. All figures show individual (gray) and aggregated (red) sensor data, for the 3 sensor units highlighted on the left plot.}
\label{fig:urbane}
\end{figure*}

Our team has been working on a 3D, urban GIS framework called \emph{Urbane} \cite{Ferreira:VAST:2015}, see Figure \ref{fig:urbane}. Urbane is an interactive tool, including a novel 3D map layer, that has been developed from the ground up by the project's data science team to take advantage of the GPU capabilities of modern computing systems. It allows for fast, potentially real-time computations, as well as the integration and visualization of multiple data streams commonly found for major cities such as NYC. In the context of SONYC, we have expanded Urbane's capabilities to include the efficient management of high-resolution, temporal data. This is achieved via a novel data structure, called the \emph{time lattice}, which allows for  fast retrieval, visualization and analysis of individual and aggregate sensor data at multiple time scales such as hours, days, weeks, and months. An example of data retrieved using this capability can be seen on the right plot in Figure \ref{fig:urbane}. Urbane and the time lattice, have been used to support the preliminary noise analysis in Section \ref{sec:mitigation}, but their applicability goes well beyond audio. 

We are currently expanding Urbane to support visual spatio-temporal queries over noise data, including the use of computational topology methods for pattern detection and retrieval. Similar tools have already proved useful in smart cities situations, such as prior collaborations between team members, the NYC Department of Transportation and the Taxi and Limousine Commission \cite{taxivis@tvcg2013,Doraiswamy:VIS:2014}.

\section{Making the Case for Data-driven Mitigation} 
\label{sec:mitigation}

We conducted a preliminary study on the validity and response of noise complaints around the Washington Square area of NYC using SONYC's current sensing and analytics infrastructure \cite{mydlarz_internoise_2017}. The study combines information mined from the log of civic complaints made to the city over the last year via its 311 system, the analysis of a subset of our own sensor data during the same period, and information gathered via interactions and site visits with inspectors from NYC's Department of Environmental Protection (DEP) tasked with enforcing the noise code.

\begin{figure*}[h]
\centering\includegraphics[width=0.8\linewidth]{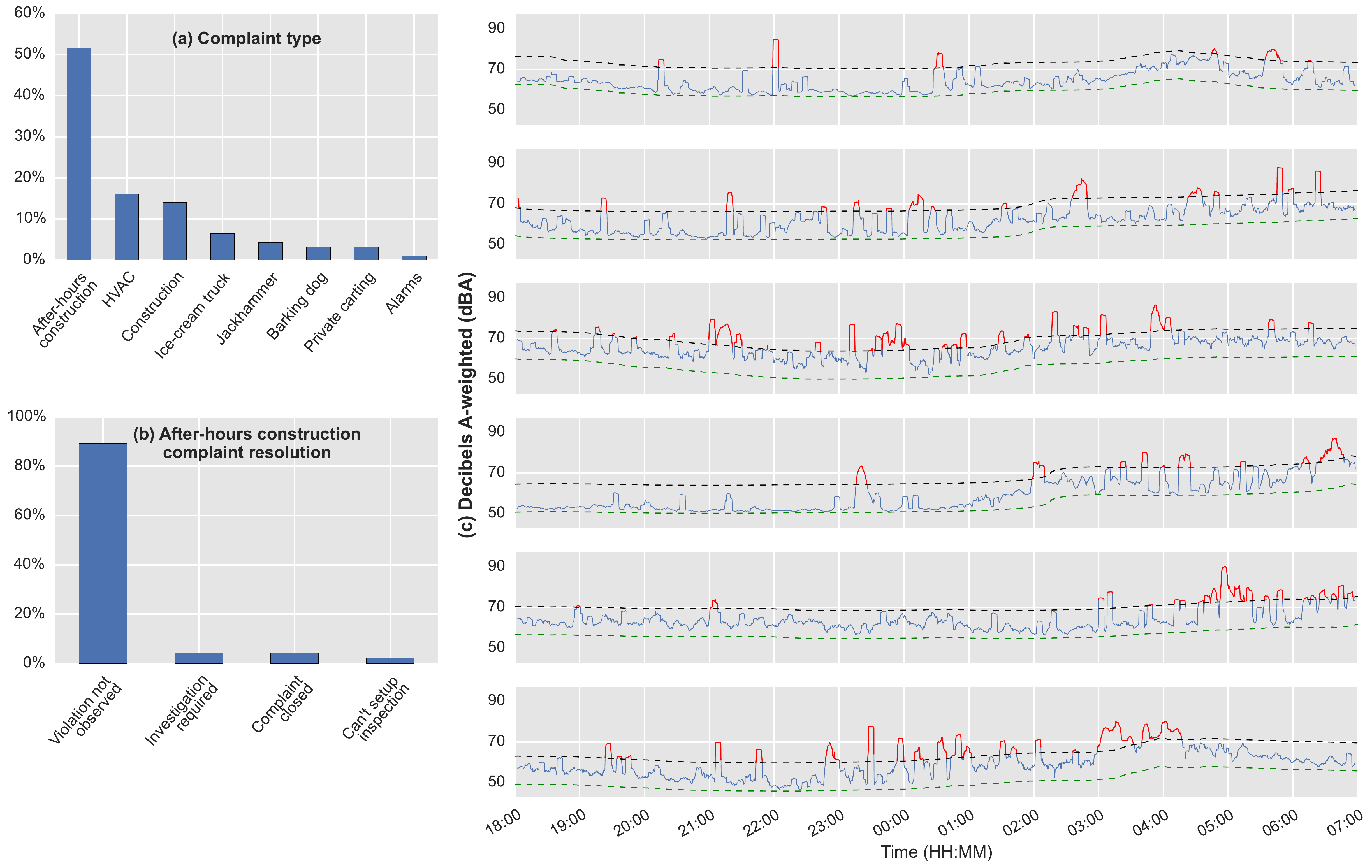}
\caption{(a) Distribution of 311 outdoor noise complaints in the focus area during the period of study. The bar graph shows a clear predominance of after-hours construction noise. (b) Distribution of complaint resolutions for after-hours construction complaints. The vast majority of complaints result in a \emph{violation not observed} status. (c) Sensor data for the after-hours period corresponding to six complaints: continuous SPL data (blue), background level (green), event detection threshold at 10dB above background level (black), potential noise code violation events (red).}
\label{fig:complaint_by_res}
\end{figure*}

For our study we chose an area with a relatively dense deployment of 17 nodes. We established a 100\,m boundary around each node and merged to form the focus area. From 311, we collected all non-duplicate noise complaints occurring within this area that were routed to the DEP while neighboring sensors were active. Note that this criterion discards, for example, complaints about noise from neighbors, which are routed to the police department, and tend to dominate the 311 log. The breakdown of selected complaint types can be observed in Figure \ref{fig:complaint_by_res}(a).

Over an 11-month period, 51\% of all noise complaints in the focus area were related to after-hours construction activity (between 6PM--7AM), 3 times the amount of the next category. Note that combining all construction-related complaints adds up to 70\% of this sample, highlighting the importance of this issue.

Figure \ref{fig:complaint_by_res}(c) shows SPL values (blue line), at a 5-minute resolution, for the after-hours period during or immediately preceding a subset of the complaints. Dotted green lines correspond to background levels, computed as the moving average of SPL measurements within a 2-hour window. Dotted black lines correspond to SPL values 10dB above the background, the threshold defined by the city's noise code to indicate potential violations. Finally, we can detect events (marked in red), where instantaneous SPL measurements are above the threshold. 
Our analysis resulted in the detection of 324 such events. 
We classified those by noise source, and determined that 76\% (246) were related to construction as follows: jackhammering (223), compressor engine (16) and metallic banging/scraping (7); with the remainder corresponding to non-construction sources, predominantly sirens and other traffic noise. Our analysis shows that for 94\% of all after-hours construction complaints, we can find quantitative evidence in our sensor data of a potential violation.

How does this evidence stack against the enforcement record for those complaints? Citizen complaints submitted via 311 and routed to the DEP trigger an inspection, with the city open data including information about how each complaint is resolved. For all complaints in this study, 
78\% resulted in a \textit{``No violation could be observed''} status, and only 2\% in a violation ticket being issued. Figure \ref{fig:complaint_by_res}(b) shows that in the specific case of after-hours construction noise, no violation could be observed in 89\% of all cases, and none of the inspections resulted in a violation ticket being issued. 

There are multiple explanations for the significant gap between the evidence collected by the sensor network and the results of the inspections. For example, we can speculate that the mismatch is partly due to the delay in the city's response to complaints, 4-5 days in average, which is too high for phenomena that are both transient and trace-less. Another factor is the conspicuousness of the inspection crew, which in itself modifies the behavior of potentially offending sources, as we observed during site visits with the DEP. Moreover, under some circumstances the city government grants special, after-hours construction permits under the assumption of minimal noise impact as defined by the noise code. Thus, it is possible that some of this after-hours activity results from those permits. 
Unfortunately after-hours construction permit data is not readily available, and we are currently exploring alternative mechanisms for its collection.  

In all cases, we argue that the SONYC sensing and analytical framework can address the shortcomings of current monitoring and enforcement mechanisms by providing hard data to: (a) quantify the actual impact of after-hours construction permits on the acoustic environment (and thus the city inhabitants); (b) provide historical data that can validate complaints and thus support inspection efforts in an inconspicuous and continuous basis; and (c) develop novel, data-driven strategies for the efficient allocation of inspection crews in space and time using the same tools from operations research that optimize routes for delivery trucks and taxis. It is worth noting that, even though our preliminary study focused on validating 311 complaints, SONYC can be used to gain insight beyond the complaint data, allowing us to understand the extent and type of unreported noise events, to identify biases in complaint behavior, and to accurately measure the level of noise pollution in the environment.

\section{Looking Forward}

The SONYC project has just completed the first of five years of its research and development agenda. The current focus of the project is squarely on the development and deployment of the intelligent sensing infrastructure, but as the work progresses that focus will progressively shift towards analytics and mitigation, in collaboration with city agencies and other stakeholders. Here are some areas of future work. 

\noindent \underline{Low-power mesh sensor network:} to support deployment of sensors at significant distances from Wi-Fi or other communication infrastructure and at locations lacking easy access to wired power, 
we are developing a second generation of the sensor node that is mesh enabled and battery/solar powered. Each sensor node will itself serve as a router in a low-power multi-hop wireless network in the 915MHz band, using FCC-compatible cognitive radio techniques over relatively long links and energy-efficient multi-channel routing for communicating to and from infrastructure-connected base stations. Our sensor design will further reduce power consumption for the multi-label noise classification, by leveraging heterogeneous processors for duty-cycled/event-driven hierarchical computing.  Specifically, our sensor node is based on a low power System-on-Chip, the Ineda i7~\footnote{http://inedasystems.com/wearables.php}, 
for which we are redesigning ``mote-scale'' computation techniques originally developed for single micro-controller devices to now support heterogeneous processor-specific operating systems via hardware virtualization.

\noindent \underline{Modeling:} the combination of noise data collected by sensors and citizens will be necessarily sparse in space and time. In order to perform meaningful analyses and help inform decisions by city agencies, it is essential to compensate for this sparseness. Several open data sets are available that could, either directly or indirectly, provide information on the noise levels in the city: the locations of restaurants, night clubs and tourist hot spots indicate areas where social noise sources are likely, while social media data streams can be used to understand the temporal dynamics of crowd behavior. Likewise, multiple data streams about, e.g., taxi, buses and aircrafts, can provide indirect information on traffic-based noise levels. We plan to develop noise models that use spatio-temporal covariance to predict unseen acoustic responses using a combination of sensor and open data. We will also explore combinations of this data-driven modeling, with physical models that exploit the 3D geometry of the city, sound type and localization cues from sensors and 311, and basic sound propagation principles. We expect that through a combination of techniques from data mining, statistics, and acoustics, as well as our significant expertise in developing models suitable for GPU implementation using ray casting queries in the context of computer graphics, we can pioneer the generation of accurate, dynamic and 3D urban noise maps in real time.

\noindent \underline{Citizen Science and civic participation:} the role of humans in SONYC is not limited to annotating sound. In addition to the fixed sensors located in various parts of New York City, a SONYC mobile platform is expected to enable citizens to record and annotate sounds in situ, view existing data contributed and analyzed by others, and contact authorities about noise-related concerns. A mobile platform will allow users to leverage slices taken from this rich set of data to describe their concerns, and support them with evidence, as they approach city authorities, regulators, and policy makers. Citizens will  not only be more informed and more engaged with their environment, but also better equipped in voicing their concerns in effective ways as they interact with authorities.

\smallskip

SONYC is a smart cities, next-generation application of cyber-physical systems. Its development calls for innovation in various fields of computing and engineering, including sensor networks, machine learning, human-computer interaction, citizen science and data science. Furthermore, our technology can support novel scholarly work on noise pollution in public health, public policy, environmental psychology and economics. But the project is far from a purely scholarly endeavor. By seeking to improve urban noise mitigation, a critical quality-of-life issue, SONYC has direct potential to benefit urban citizens around the world. Our agenda calls for SONYC to be deployed, tested and used in real-world conditions, the outcome potentially a model that can be scaled and replicated across the US and beyond.

\begin{acks}
This work is partially supported by the National Science Foundation (Award \# 1544753), NYU's Center for Urban Science and Progress, NYU Tandon School of Engineering, and the Translational Data Analytics Institute at OSU.
\end{acks}

\bibliographystyle{ACM-Reference-Format}
\bibliography{bib/prior,bib/analysis,bib/vis,bib/citsci,bib/proposal,bib/sensing}

\end{document}